\documentclass[aps,onecolumn,preprintnumbers,showpacs,showkeys,nofootinbib,
superscriptaddress  
]{revtex4}

\usepackage{epsfig}
\usepackage{amssymb,amsmath,amsfonts,amsthm,graphicx,psfrag}
\newcommand{\beq}{\begin{equation}}
\newcommand{\eeq}{\end{equation}}
\newcommand{\bea}{\begin{eqnarray}}
\newcommand{\eea}{\end{eqnarray}}

\newcommand{\tr}{\operatorname{Tr}}

\begin{document}
\preprint{}

\title{
Study of shear viscosity of $SU(2)$-gluodynamics within lattice simulation
}

\author{N.~Yu.~Astrakhantsev}
\email[]{nikita.astrakhantsev@itep.ru}
\affiliation{Institute for Theoretical and Experimental Physics, Moscow, 117218 Russia\\
and Moscow Institute of Physics and Technology, Dolgoprudny, 141700 Russia}

\author{V.~V.~Braguta}
\email[]{braguta@itep.ru}
\affiliation{Institute for High Energy Physics NRC "Kurchatov Institute", Protvino, 142281 Russian Federation\\
Far Eastern Federal University, School of Biomedicine, 690950 Vladivostok, Russia \\
and Institute for Theoretical and Experimental Physics, Moscow, 117218 Russia \\ 
and National Research Nuclear University MEPhI (Moscow Engineering Physics Institute), 115409, Russia, Moscow, Kashirskoe highway, 31
}

\author{A.~Yu.~Kotov}
\email[]{kotov@itep.ru}
\affiliation{Institute for Theoretical and Experimental Physics, Moscow, 117218 Russia}

\begin{abstract}
This paper is devoted to the study of two-point correlation function of the energy-momentum tensor
$\langle T_{12} T_{12} \rangle$ for $SU(2)$-gluodynamics within lattice simulation of QCD. Using multilevel algorithm we carried out 
the measurement of the correlation function at the temperature $T/T_c \simeq 1.2$. It is shown that 
lattice data can be described by spectral functions which interpolate between hydrodynamics at 
low frequencies and asymptotic freedom at high frequencies. The results of the study of 
spectral functions allowed us to estimate the ratio of shear viscosity to the entropy density 
${\eta}/{s} = 0.134 \pm 0.057$.
\end{abstract}



\maketitle

\subsection*{Introduction}

One of the most important result obtained at RHIC experiment is the measurement of the elliptic flow of final particles\cite{Adams:2005dq,Adcox:2004mh}. 
The value of the elliptic flow measured at RHIC can be explained within the approach based on hydrodynamics \cite{Kolb:2000fha,Huovinen:2001cy,Teaney:2000cw}, 
if one assumes that quark-gluon plasma (QGP) obtained after the collision is almost superfluid. In particular, numerical simulations of the relativistic 
liquid showed \cite{Teaney:2009qa} that the upper limit on the ratio of shear viscosity to entropy density is $\eta/s \le 0.4$.
Preferred range for the ratio is $\eta / s = (1 \leftrightarrow 3) \times 1 / 4 \pi$, which is very close to the result of N = 4 Super Yang Mills (SYM)
theory at strong coupling $\eta / s = 1 / 4 \pi$\cite{Policastro:2001yc}. 
From these facts one can conclude that theoretical prediction of the ratio $\eta / s$ is a very interesting and important problem. 

It is known that QGP is strongly interacting system and today there are no analytical approaches which allow to study such 
systems without additional assumptions. For this reason one of the main approaches which can be used to study the properties of QGP
and which is based on the first principles is lattice simulation of QCD. 

Lattice simulation of QCD aimed at the calculation of shear viscosity of the $SU(3)$-gluodynamics was carried out 
in papers \cite{Karsch:1986cq, Nakamura:2004sy, Meyer:2007ic, Meyer:2009jp}. Despite rather large uncertainties one can 
state that the ratios $\eta/s$ obtained in papers \cite{Meyer:2007ic, Meyer:2009jp} are close to the N = 4 SYM prediction 
$\eta / s \sim 1 / 4 \pi$. The calculation of shear viscosity in QCD with dynamical quarks is still a challenging problem. 

The closeness of the SU(3)-gluodynamics ratio $\eta/s$ to the N = 4 SYM prediction allows us to ask 
the question whether this property is the property of the SU(3)-gluodynamics or it is common to all non-abelian 
gauge theories. To address this question in this paper we are going to perform the calculation of shear viscosity 
of $SU(2)$-gluodynamics within lattice simulation.  

We have already performed the first measurement of shear viscosity of the $SU(2)$-gluodynamics 
in \cite{Braguta:2013sqa}. In this paper we extend our investigation to larger lattice size and apply different method 
to determine numerical value of the ratio $\eta / s$.

\subsection*{Details of the calculation}

Shear viscosity is related to the Euclidean correlation function of the energy-momentum tensor 
$T_{\mu \nu} = \frac 1 4 \delta_{\mu \nu} F_{\alpha \beta}^a F_{\alpha \beta}^a - F_{\mu \alpha }^a F_{\nu \alpha}^a$ (here we omitted for simplicity trace anomaly):
\begin{equation}\begin{split}
	C(x_0)=T^{-5} \int d^3{\textbf x}\langle T_{12}(0)T_{12}(x_0,{\textbf x})\rangle,
\label{correlator}
\end{split}\end{equation}
where $T$ is the temperature of the system.
The correlation function (\ref{correlator}) can be written in terms of the spectral function $\rho(\omega)$ as follows
\begin{equation}\begin{split}
	C(x_0)=T^{-5} \int_{0}^{\infty}\rho(\omega)\frac{\cosh \omega(\frac1{2T}-x_0)}{\sinh\frac{\omega }{2T}} d\omega.
\label{spectr_corr}
\end{split}\end{equation}

The spectral function contains a lot of important information about the properties of medium.
In particular, to find shear viscosity from spectral function one uses the Kubo formula \cite{Kubo:1957mj} 
\begin{equation}\begin{split}
	\eta=\pi\lim\limits_{\omega\to0} \frac{\rho(\omega)}{\omega},
\end{split}\end{equation}
Lattice calculation of shear viscosity can be divided into two parts. The first part is the measurement 
of the correlation function $C(x_0)$ with sufficient accuracy. 
This part of the calculation requires large computational resources but for the gluodynamics the accuracy of the correlator can be dramatically improved with the help of the two-level algorithm \cite{Meyer:2002cd}.
The second part is the determination 
of the spectral function $\rho(\omega)$ from the correlation function $C(x_0)$. The last part of 
the calculation is probably the most complicated, since one should determine continuous spectral
 function $\rho(\omega)$ from integral equation (\ref{spectr_corr})
for the set of $O(10)$ values of the function $C(x_0)$ measured in lattice simulation. 

To approach the solution of the integral equation (\ref{spectr_corr}) one should take into account the properties
of the spectral function. Important properties of the spectral function are positivity: $\rho(\omega)/\omega \ge 0$ and oddness: $\rho(-\omega) = - \rho( \omega)$. 
It is also important to write the expression for the spectral function 
at the leading-order approximation in strong coupling constant \cite{Meyer:2008gt}
\begin{equation}
\rho^{LO} (\omega) = \frac 1 {10} \frac {d_A} {(4 \pi)^2} \frac {\omega^4} {\tanh (\frac {\omega} {4T})} + \biggl ( \frac {2 \pi} {15} \biggr )^2 d_A T^4 ~ \omega \delta(\omega),
\label{rho_tree_level}
\end{equation}
where $d_A=N_c^2-1=3$ for the SU(2)-gluodynamics. One also knows next-to-leading order expression for the 
spectral function at large $\omega$ \cite{Kataev:1981gr}
\begin{equation}
\lim_{\omega \to \infty} \rho^{NLO} (\omega) = \frac 1 {10} \frac {d_A} {(4 \pi)^2} \omega^4 \biggr (1 - \frac {5 \alpha_s N_c} {9 \pi}  \biggl )
\label{rho_NLO}
\end{equation}
It should be noted here that at large $\omega$ the spectral function scales as $\rho(\omega) \sim \omega^4$,
what leads to a large perturbative contribution to the correlation function for all values of Euclidean time $x_0$.
Calculation shows that even at the $x_0 = 1/(2T)$ the tree level contribution is $\sim 85\%$ of the total value of the 
correlation function. Note also that large $\omega$ behaviour of the spectral function 
leads to a fast decrease of the correlation function $C(x_0) \sim 1/x_0^5$ for small $x_0$. 
For this reason the signal/noise ratio for the $C(x_0)$ is small at $x_0 \gg a$
and lattice measurement of the correlation function at $x_0 \sim 1/(2T)$ becomes computationally very expensive.

In numerical simulation we use Wilson gauge action for the SU(2)-gluodynamics
\begin{equation}\begin{split}
	S_g=\beta\sum_{x,\mu<\nu}\left(1-\frac1{2}\tr U_{\mu,\nu}(x)\right),
\end{split}\end{equation}
where $U_{\mu,\nu}(x)$ is the product of the link variables along elementary rectangular $(\mu, \nu)$, which starts at $x$.

For the tensor $F_{\mu\nu}$ we use the clover discretization scheme:
\begin{equation}\begin{split}
	F^{(clov)}_{\mu\nu}(x)=\frac1{4iga^2}(V_{\mu,\nu}(x)+V_{\nu,-\mu}(x) \\
	+ V_{-\mu,-\nu}(x)+V_{-\nu,\mu}(x) )\\
	V_{\mu,\nu}(x)=\frac12(U_{\mu,\nu}(x)-U_{\nu,\mu}(x)).
\end{split}
\end{equation}
It causes no difficulties to build energy-momentum tensor having expression for  the tensor $F_{\mu\nu}$. 

To calculate shear viscosity one should measure the correlation function (\ref{correlator}). To carry out this measurement 
we use two-level algorithm described in \cite{Meyer:2002cd}. This algorithm significantly improves the speed of the calculations. Note also
that instead of the correlation function $\langle T_{12}(x)T_{12}(y)\rangle$ in this paper we measure the correlation function 
$\frac12(\langle T_{11}(x)T_{11}(y)\rangle-\langle T_{11}(x)T_{22}(y)\rangle)$. Both correlation functions are equal in 
the continuum limit\cite{Karsch:1986cq}. 

It has become conventional to present the value of shear viscosity as a the ratio viscosity-to-entropy density $\eta/s$. For homogeneous systems the entropy density $s$ can be expressed as $s=\frac{\epsilon+p}T$, where $\epsilon$ is the energy density and $p$ is the pressure. These thermodynamic quantities were 
measured with the method described in \cite{Engels:1999tk}. 	

Energy-momentum tensor in continuum theory is a set of Noether currents which are related to the
translation invariance of the action.  In lattice formulation of field theory continuum translation invariance does not exist 
and renormalization for energy-momentum tensor is required. 
For the correlation function considered in this paper the renormalization is multiplicative \cite{Meyer:2011gj}. 
Renormalization factors depend on the discretization scheme. For instance, for the diagonal component of 
$T_{\mu\nu}$ (when $\mu=\nu$) and the plaquette-based discretization of $T_{\mu\nu}$: 
$T_{\mu\mu}=\frac{2}{a^4g^2}\left(-\sum\limits_{\nu\ne\mu}\tr U_{\mu,\nu}(x)+\sum\limits_{\nu,\sigma\ne\mu,\sigma>\nu}\tr U_{\sigma,\nu}(x)\right)$ the renormalization factors are related to the anisotropy coefficients \cite{Engels:1981qx,Meyer:2007fc}: $T^{(ren)}_{\mu\nu}=Z^{(plaq)}T_{\mu\nu}^{(plaq)}$, $Z^{(plaq)}=1-\frac12g_0^2(c_{\sigma}-c_{\tau})$, where $c_{\sigma}$ and $c_{\tau}$ are defined in \cite{Engels:1999tk} and for $SU(2)$ are computed on the lattice in \cite{Engels:1994xj}.
 
 Using the renormalization factors for the plaquette-based discretization of $T_{00}$, we can find the renormalization factors for the clover discretization simply by fitting the vacuum expectation values of the renormalized $T_{00}$: 
$Z^{(plaq)}\langle T_{00}^{(plaq)}\rangle=Z^{(clov)}\langle T_{00}^{(clov)}\rangle$.

\section*{Numerical results}

\begin{figure}[t]
\begin{center}
\includegraphics[scale=1.]{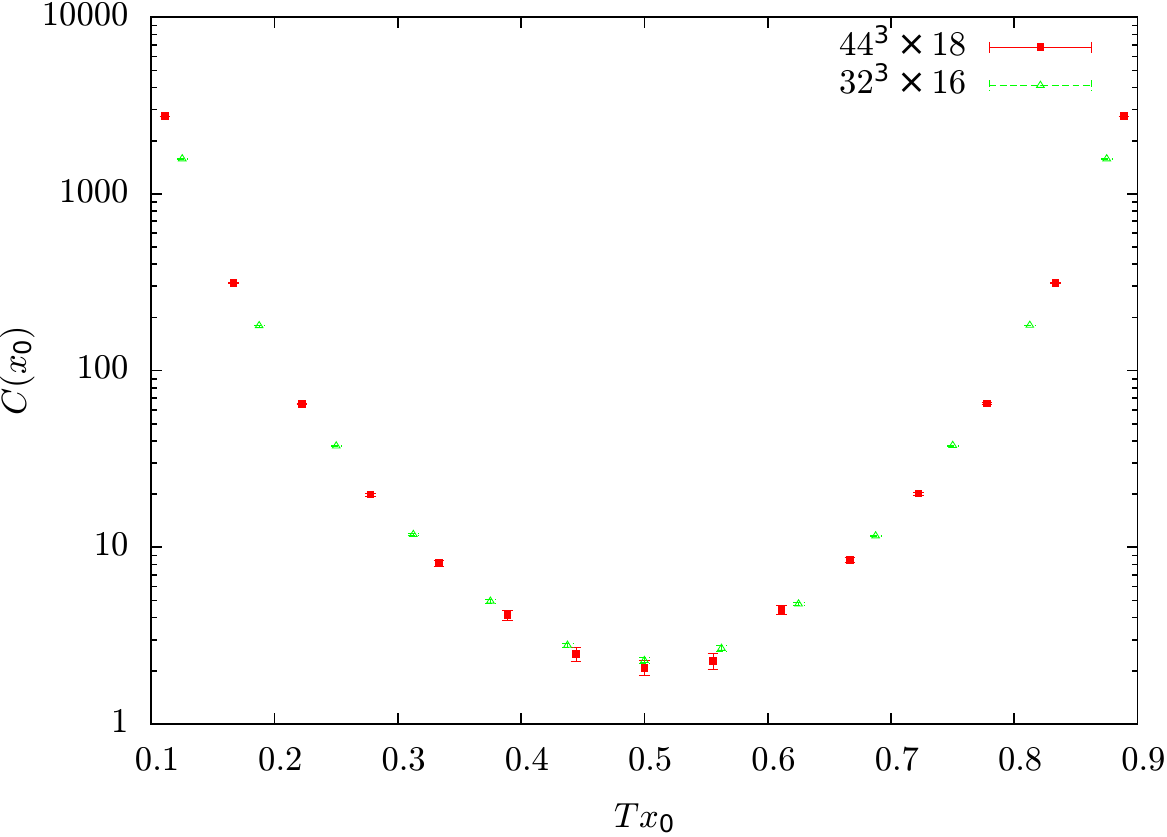}
\caption{Renormalized correlation functions $C(x_0)$
as a function of Euclidean time $x_0$ measured at lattices  $16\times32^3$ and $18\times44^3$}
\label{fig:corr_func}
\end{center}
\end{figure}

  We carried out the measurements of the correlation function $C(x_0)$ at lattices $16\times32^3$ with $\beta=2.81$ and $18\times44^3$ with $\beta=2.85$. 
Both lattices correspond to deconfinement phase with the temperature $T/T_c \simeq 1.2$ but different physical volumes and lattice spacings.  
In Fig.\ref{fig:corr_func} we plot renormalized correlation functions (\ref{correlator})
as a function of Euclidean time $x_0$ for the lattices $16\times32^3$ and $18\times44^3$. Two-level algorithm allowed us 
to reach accuracy $\sim 3\%$ for the lattice $16\times32^3$ and $\sim 10\%$ for the lattice $18\times44^3$ at point $T x_0 = 0.5$. For the other 
points the accuracy is much better. 

Visually the set of values of the correlation functions measured at different lattices lies on one curve. The values for the correlation functions 
measured at close points in the region $T x_0 \sim 0.5$ are very close to each other. Note also that the fit parameters (see below) of both 
correlation functions are very close to each other. These facts allow us to expect that finite lattice size and finite volume 
effects are not very important in our calculation. 

\begin{figure}[t]
\begin{center}
\includegraphics[scale=1.]{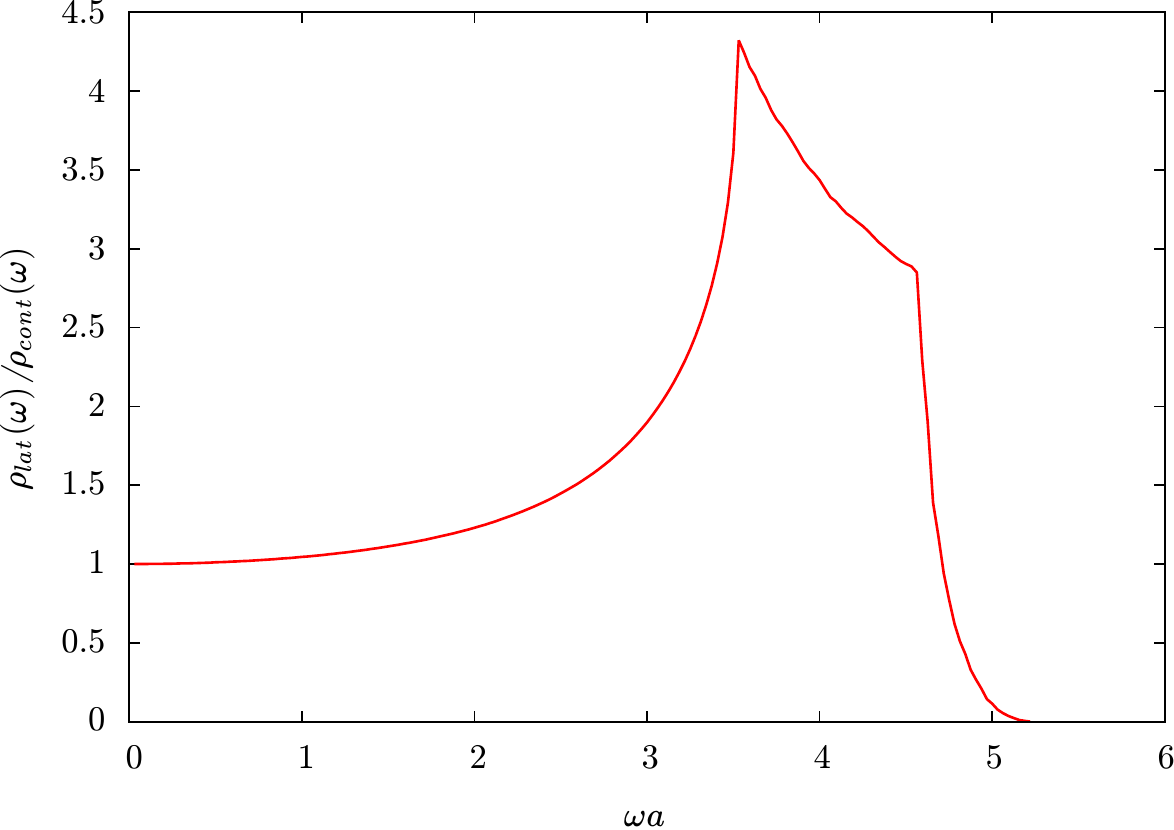}
\caption{The ratio $\rho_{lat}(\omega)/\rho_{cont}(\omega)$ as a function of $\omega a$}
\label{fig:rholat}
\end{center}
\end{figure}

Unfortunately the accuracy of our results and number of points where the correlation function (\ref{correlator}) was measured 
do not allow us to use Maximal Entropy Method to determine the spectral function $\rho(\omega)$. For this reason we use 
physically motivated ansatz for the spectral function with unknown parameters which will be determined through the fitting procedure. 
Probably the simplest formula for the spectral function inspired by QCD sum rules\cite{Shifman:1978bx} can be built 
if we join hydrodynamical behaviour at small frequencies with asymptotic freedom at large frequencies
\footnote{Note that the frequency $\omega$ is measured in physical units.}
\begin{equation}
\rho_1(\omega) = B T^3~\omega ~ \theta(\omega_0 - \omega) + A  \rho_{lat} ( \omega )~ \theta(\omega-\omega_0).
\label{rho1}
\end{equation}
In last formula $\rho_{lat}(\omega)$ is a tree level lattice expression for the spectral function
calculated for the correlation function 
$\sim \frac12(\langle T_{11}(x)T_{11}(y)\rangle-\langle T_{11}(x)T_{22}(y)\rangle)$ with clover discretization
of the tensor $F_{\mu \nu}$ at lattice with fixed $L_t$ and $L_s \to \infty$. Although the calculation of the 
$\rho_{lat}(\omega)$ can be easily performed using formulas from paper \cite{Meyer:2009vj}, the 
resulting expression is very cumbersome. For this reason instead of the explicit expression for the $\rho_{lat}$,
in Fig.\ref{fig:rholat}  we plot the ratio $\rho_{lat}(\omega)/\rho_{cont}(\omega)$ for the lattice $L_t=16$. 

\begin{figure}[t]
\begin{center}
\includegraphics[scale=1.]{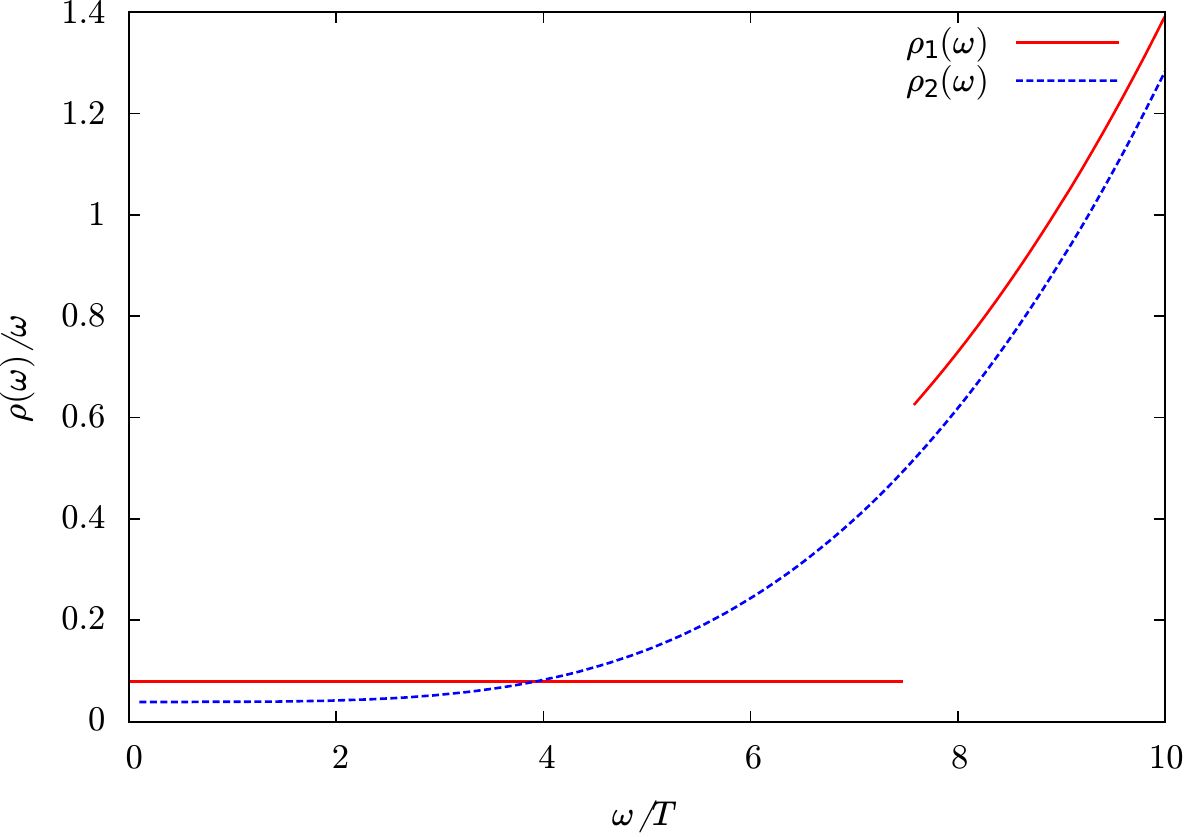}
\caption{Spectral functions $\rho_1(\omega)$ and $\rho_2(\omega)$ as the function of $\omega/T$ for the lattice $L_t=16$.}
\label{figrho12}
\end{center}
\end{figure}

The fit of the lattice data ($x_0/a \ge 2$) with the formula (\ref{spectr_corr}) with the spectral function (\ref{rho1})
gives $A = 0.723 \pm 0.002, B = 0.079 \pm 0.016, \omega_0/T = 7.5 \pm 0.5, \chi^2 / dof \simeq 1.4$  for the lattice $16\times32^3$ 
and $A = 0.703 \pm 0.003, B = 0.096 \pm 0.026, \omega_0/T = 8.3 \pm 0.7, \chi^2 / dof \simeq 0.7$ for the lattice $18\times44^3$.
It is seen that ansatz (\ref{rho1}) for the spectral function fits the lattice data very well and 
the parameters of the fit for different lattices are in a reasonable agreement with each other. 
The resulting shear viscosity is $\eta/s = 0.179 \pm 0.036$ for the lattice $16\times32^3$ and 
$\eta/s = 0.217 \pm 0.059$ for the lattice $18\times44^3$. In Fig.\ref{figrho12} we plot the spectral 
function $\rho_1(\omega)$ as a function $\omega$ for the lattice $L_t=16$.

Now few comments are in order
\begin{itemize}
\item In formula (\ref{rho1}) we used $\rho_{lat}(\omega)$ instead of the continuum tree-level expression for the $\rho(\omega)$ 
(see formula (\ref{rho_tree_level})). We believe that this allows us to take into account discretization uncertainty 
and asymptotic freedom contribution more carefully. If one puts the continuum tree-level expression to formula 
(\ref{rho1}), the $\chi^2 / dof$ of the fits will be considerably enhanced: 
$\chi^2 / dof \simeq 5.2$ for the lattice $16\times32^3$ and $\chi^2 / dof \simeq 4.5$ for the lattice $18\times44^3$.

\item As was noted above the asymptotic behaviour of the spectral function at large $\omega$ is fixed $\rho \sim \omega^4$.
However, the coefficient in front of the $\omega^4$--behaviour is modified by higher order radiative corrections (see formula (\ref{rho_NLO})). 
In order to take this effect into account in formula (\ref{rho1}) we introduced factor $A$. 
At this point our ansatz deviates from that used in papers \cite{Meyer:2007ic, Meyer:2009jp}. 
If we take $A=1$, the ansatz (\ref{rho1}) will not be able to describe our data ($\chi^2/dof > 100$ for both lattices).

\item The values of the fit parameters are physically well motivated. For instance, 
the value of the parameter $A$ is smaller than unity, what agrees with the next-to-leading order 
result (\ref{rho_NLO}). The value of the strong coupling constant at the threshold parameter $\omega_0$ 
($\omega_0 \sim 2.7$ GeV in physical units) is $\alpha_s(\omega_0) \sim 0.2-0.3$. This allows us 
to expect that perturbative expression for the spectral function is applicable for $\omega > \omega_0$.

\end{itemize}

Low frequency part of the spectral function (\ref{rho1}) is given by the first-order hydrodynamic 
expression $\sim \omega$. Comparison of the spectral functions of the energy-momentum 
tensor correlation functions obtained in N = 4 SYM \cite{Kovtun:2006pf} and the first-order hydrodynamic 
expressions allows us to expect that this approximation works well up to $\omega \leq \pi T \simeq 1$ GeV\cite{Meyer:2008sn}. 
From the other side high frequency perturbative expression for the spectral function is fixed very 
accurately and it works well for $\omega \geq \omega_0 = 2.7$ GeV. The form of the spectral function 
in the region $1~\mbox{GeV} \leq \omega \leq 2.7~\mbox{GeV}$ is not clear. Formula (\ref{rho1})
continues the first-order hydrodynamic expression to the region $1~\mbox{GeV} \leq \omega \leq 2.7~\mbox{GeV}$ (see Fig. \ref{figrho12}). 
As the result 
there is rather large discontinuity  in the spectral function (\ref{rho1}) at point $\omega_0$ 
which for the lattice $16 \times 32^3$ is $(\rho_1(\omega_0+0)-\rho_1(\omega_0-0))/\rho_1(\omega_0-0) \simeq 6.7$. 
The last fact allows us to state that the spectral function in form (\ref{rho1}) underestimates 
real spectral function in the region $1~\mbox{GeV} \leq \omega \leq 2.7~\mbox{GeV}$ and as a result the 
value of shear viscosity obtained in this fit is larger than its real value. 

As was noted a serious drawback of the spectral function $\rho(\omega)$ is its discontinuity at point $\omega=\omega_0$. 
In order to solve this problem we propose another form of the spectral function which preserves
basic ideas of the spectral function (\ref{rho1}) and takes into account the property $\rho(-\omega)=-\rho(\omega)$
\begin{equation}
\rho_2(\omega) = B T^3 \omega + A \rho_{lat}(\omega)   \tanh^{2} \biggl ( \frac {\omega} {\omega_0} \biggr ).
\label{rho2}
\end{equation}

The fit of the lattice data ($x_0/a \ge 2$) by the formula (\ref{spectr_corr}) with the spectral function (\ref{rho2})
gives $A = 0.723 \pm 0.003, B = 0.039 \pm 0.012, \omega_0/T = 5.6 \pm 0.6, \chi^2 / dof \simeq 1.2$  for the lattice $16\times32^3$ 
and $A = 0.705 \pm 0.004, B = 0.055 \pm 0.022, \omega_0/T = 6.5 \pm 0.9, \chi^2 / dof \simeq 0.6$ for the lattice $18\times44^3$.
It is seen that ansatz (\ref{rho2}) for the spectral function fits the lattice data very well 
and parameters from different lattices are in a reasonable agreement with each other. 
The resulting shear viscosity is $\eta/s = 0.088 \pm 0.027$ for the lattice $16\times32^3$ and 
$\eta/s = 0.125 \pm 0.050$ for the lattice $18\times44^3$. In Fig.\ref{figrho12} we plot the spectral 
function $\rho_2(\omega)$ as a function $\omega$ for the lattice $L_t=16$. Notice that the function (\ref{rho2}) is larger 
than (\ref{rho1}) in the region $1~\mbox{GeV} \leq \omega \leq 2.7~\mbox{GeV}$, 
as the result the value of the of shear viscosity becomes smaller. Note also that the value of shear 
viscosity obtained at the lattice $16\times32^3$ is very close to the N = 4 SYM prediction $\eta/s = 1/4\pi \simeq 0.080$.

In order to investigate the systematic errors, caused by the usage of the ansatz (8),(9)
we studied other fit functions. Instead of the $\tanh^{2} ( \omega/\omega_0)$ in 
formula (\ref{rho2})
one can use, for example, any power of this function $\sim \tanh^{2k} ( \omega/\omega_0)$ or 
linear combination $\sim \sum_k A_k \tanh^{2k} ( \omega/\omega_0)$. 
\footnote{ It should be noted here that in the region $1~\mbox{GeV} \leq \omega \leq 2.7~\mbox{GeV}$ any function 
can be approximated by this linear approximation with good accuracy.}
We found, that in these cases 
the fits are also good and the resulting viscosities are greater than that for the 
spectral function  (\ref{rho2}) and smaller than 
that for the spectral function (\ref{rho1}).

As the result of this paper we take the value 
\begin{equation}
\frac {\eta} {s} = 0.134 \pm 0.034 \pm 0.046.
\end{equation} 
The central value is the 
average between the values of shear viscosity obtained with the spectral functions in the form (\ref{rho1})
and (\ref{rho2}) from the fit of lattice data measured at the lattice $16\times32^3$. The first 
uncertainty is due to the fitting procedure which is typically $25 \%$. The second 
uncertainty is due to the unknown model of the spectral function. We estimated this uncertainty 
assuming that the central value of shear viscosity can vary between central values of the 
fits (\ref{rho2}) and (\ref{rho1}). Actually there are a lot of different sources of uncertainties
of our result but they are much smaller than statistical and spectral function model uncertainties. 

\section*{Conclusion}

In this paper we studied the energy-momentum tensor correlation function $\langle T_{12}(0) T_{12}(x) \rangle$
for $SU(2)$-gluodynamics using lattice simulation of QCD. We carried out the measurements of this correlation function 
at lattices $16\times32^3$ with $\beta=2.81$ and $18\times44^3$ with $\beta=2.85$. 
Both lattices correspond to deconfinement phase with the temperature $T/T_c \simeq 1.2$ but different physical volumes and lattice spacings.  
In order to enhance the accuracy of the calculation we used two-level algorithm which allowed us 
to reach accuracy $\sim 3\%$ for the lattice $16\times32^3$ and $\sim 10\%$ for the lattice $18\times44^3$ at point $x_0 = 1/2T$. 
For the other points the accuracy is much better. 

Using lattice data for the correlation function we tried to study the spectral function. It was shown that 
physically motivated anzatz which joins the first-order hydrodynamical behaviour at small frequencies with asymptotic freedom 
at large frequencies fits our data very well for both lattices. We also studied other forms 
of the spectral functions which interpolate between hydrodynamics and asymptotic freedom. 

The results of the study of spectral functions allowed us to estimate the ratio of shear viscosity to the entropy density
${\eta}/{s} = 0.134 \pm 0.057$. This value is in agreement with our previous finding ${\eta}/{s} = 0.111 \pm 0.032$ \cite{Braguta:2013sqa} 
obtained at small lattice with the estimation of statistical uncertainty only. 
Note also that within the uncertainty of the calculation the value obtained in this paper 
agrees with the value measured in $SU(3)$-gluodynamics: ${\eta}/{s} = 0.102 \pm 0.056$
measured at temperature $T/T_c=1.24$\cite{Meyer:2007ic}. 

The values of the ratio $\eta/s$ for the $SU(2)$ and $SU(3)$-gluodynamics are very close to the prediction of the N = 4 SYM  
theory at strong coupling: $\eta / s = 1 / 4 \pi$. So one can state that both theories belong to strongly correlated  
systems and probably the closeness of the ratio $\eta/s$ to the ratio in N = 4 SYM theory is the property on 
non-abelian gauge theories.

\section*{Acknowledgements}

The simulations were performed at ITEP supercomputer.
The work was supported by Far Eastern Federal University, Dynasty foundation, 
by RFBR grants 14-02-01185-a, 15-02-07596-a, 15-32-21117 by a grant of the president of the RF, MD-3215.2014.2,
and by a grant of the FAIR-Russia Research Center.

\end{document}